\documentclass[aps,pra,amsmath,amsfonts,floatfix,superscriptaddress,notitlepage,nofootinbib,twocolumns]{revtex4-1}
\usepackage{graphicx,graphics}
\usepackage[T1]{fontenc}
\usepackage{epstopdf}
\usepackage{amsmath}
\usepackage{amssymb}
\usepackage[colorlinks=true]{hyperref}
\usepackage{etoolbox}

\let\m=\mu      \let\eps=\epsilon

\def\to{\rightarrow}

\def\ha{\hat a}
\def\had{\hat a^\dagger}

\begin{document}

\title{Hubbard-Stratonovich transformation and consistent ordering in the coherent state path integral: insights from stochastic calculus
}
\author{Adam Ran\c con}
\affiliation{Universit\'e de Lille, CNRS, UMR 8523 -- PhLAM -- Laboratoire de Physique des Lasers Atomes et Mol\'ecules, F-59000 Lille, France}

\begin{abstract}
Recently, doubts have been cast on the validity of the continuous-time coherent state path integral. This has led to controversies regarding the correct way of performing calculations with path integrals, and to several alternative definitions of what should be their continuous limit. Furthermore, the issue of a supposedly proper ordering of the Hamiltonian operator, entangled with the continuous-time limit, has led to considerable confusion in the literature. Since coherent state path integrals are at the basis of the modern formulation of many-body quantum theory, it should be laid on solid foundations.

Here, we show that the issues raised above are coming from the illegitimate use of the (standard) rules of calculus, which are not necessarily valid in path integrals. This is well known in the context of stochastic equations, in particular in their path integral formulation. This insight allows for solving these issues and addressing the correspondence between the various orderings at the level of the path integral. We also use this opportunity to address the proper calculation of a functional determinant in the presence of a Hubbard-Stratonovich field, which shares in the controversies.
\end{abstract}
\maketitle

\section{Introduction}

Functional integrals are at the basis of the modern formulation of quantum statistical physics. Starting from the second quantized formulation of the many-body problem, one is naturally led to path integrals over coherent states, generalizing Feynman's path integral. As in Feynman's approach, the coherent state path integral (CSPI) is obtained by taking the continuous (imaginary) time limit $N\to \infty$ of a discretized version (with a finite number of time slices $N$). During this construction of the CSPI, to a given Hamiltonian operator $\hat H$, one associates an action $S_s$ which will, in general, depend on the discretization scheme $s$. While the standard version of the CSPI uses normal ordered operators, anti-normal ordering has also been introduced, leading to another discretization~\cite{KlauderBook}.

The discretized CSPI is well defined and gives the exact partition function with a controllable error, typically of order $\epsilon=\beta/N$, with $\beta$ the inverse temperature. In particular, the results are independent of the discretization scheme, as they should be.
Troubles arise in the continuous-time limit if one is not careful. Recently, Wilson and Galitski (WG) have called into question the validity of the continuous-time CSPI in the case of two simple models~\cite{Wilson2011}. By performing, in their opinion, ``exact'' calculations directly in continuous-time, they did not get the results obtained via the operator formalism (which are correct beyond any doubts). This difference was then ``fixed'' by modifying the corresponding action as if obtained by a different ordering of $\hat H$, not consistent with the underlying discretization.
This has led to a number of controversies regarding the proper way of performing the calculations~\cite{Kordas2014,Yanay2015,Taniguchi2017,Kochetov2019,Kordas2019}, and to some supposedly superior definitions of the continuous-time CSPI~\cite{Kordas2014,Bruckmann2018}.

To solve these controversies, we will use insights from the theory of stochastic differential equations, and their connection to (stochastic) path integrals. Indeed, it is well understood in this context how to deal with the subtleties of changes of discretization or of non-linear changes of variables. When performing these operations in stochastic differential equations,  one needs to devise ``substitution rules''  to take care of powers of the stochastic noise that are then generated, and that should naively vanish for smooth functions in the continuous-time limit \cite{Oksendal2010}.
Furthermore, different discretizations of the Langevin equation with multiplicative noise (describing the same microscopic dynamics) give rise to different actions in the continuous-time limit. However, when changing discretization, the substitution rules are not the same in the Langevin equation, and in the corresponding path integral. This is also true for non-linear change of variables in the path integral, where for instance the  chain rule of ordinary calculus does not apply. This has been understood properly only recently in the context of stochastic path integrals \cite{Cugliandolo2017}, and a specific discretization scheme has just been found that allows for the blind use of ordinary calculus (up to errors that disappear in the limit $N \to\infty$) \cite{Cugliandolo2018}.

While the strong connection between stochastic calculus and Feynman's path integral has been known for quite some time \cite{McLaughlin1971}, it does not seem to be the case for the CSPI. The reason might be that specific replacement rules had yet to be devised compared to Feynman's and stochastic path integrals. Here we present such rules for changes of discretization, and we take this opportunity to settle the above controversies. In Section \ref{exactBH}, we revisit the recent controversies and show that the standard continuous-time CSPI is unambiguously correct if handled properly. This is done using a Hubbard-Stratonovich (HS) transformation in the path integral. We show that the HS field behaves like a stochastic white noise and that therefore, evaluating functional determinants involving HS fields is a subtle problem. Using the insights of stochastic calculus, we can correct some errors made in the recent literature \cite{Galitski2011,Taniguchi2017,Kordas2019}.

In Section \ref{ordering}, we show how changing the discretization scheme $s$ at the level of the path integral transforms the action in the continuous-time limit. In fact, we demonstrate that each discretization scheme $s$ (to be properly defined below) corresponds to an $s$-ordering of the Hamiltonian operator (as defined by Cahill and Glauber \cite{Cahill1969}). All these different actions nevertheless give the same (exact) results if manipulated correctly.
Finally, in Section \ref{discuss} we touch upon the case of non-linear transformations, which is at the root of the discrepancy found by Wilson and Galitski, and discuss some open problems.

\section{Coherent state path integral and Hubbard-Stratonovich transformations \label{exactBH}}
\subsection{Discretized and continuous-time CSPI for the single site Bose-Hubbard model}

We address the validity of the CSPI in the simple case of the single site Bose-Hubbard model studied in Ref.~\onlinecite{Wilson2011}, defined by the Hamiltonian operator
\begin{equation}
\hat H= -\mu \,\hat n+\frac{U}{2}\hat n(\hat n-1),
\end{equation}
with $\hat n=\had\ha$ is the number operator, and $\ha$ and $\had$ are bosonic creation and annihilation operators, $[\ha,\had]=1$. The corresponding partition function is trivially
\begin{equation}
Z={\rm Tr}\left(e^{-\beta \hat H}\right)=\sum_{n=0}^\infty e^{-\beta\left(-\mu n +\frac{U}{2} n( n-1)\right)}.
\label{eq_exactZ}
\end{equation}
The corresponding (standard) discretized CSPI reads
\begin{equation}
Z_{1,N}=\int \prod_{k=1}^{N}\frac{d\psi_k d\psi_k^*}{2i\pi} e^{-\sum_{k=1}^N\left(\psi_k^*(\psi_k-\psi_{k-1})+\epsilon H_1(\psi_k^*,\psi_{k-1})\right)},
\label{eq_BH}
\end{equation}
where $\frac{d\psi_k d\psi_k^*}{2i\pi}=\frac{d {\rm Re}\psi_kd {\rm Im}\psi_k}{\pi}$ and the identification $\psi_0^{(*)}=\psi_N^{(*)}$ is assumed. This expression is obtain by writing $e^{-\beta \hat H}=\prod_{k=1}^N(1-\epsilon\hat H)+\mathcal{O}(\epsilon^2)$ and inserting $N-1$ times the resolution of the identity $1=\int \frac{d\psi d\psi^*}{2i\pi}e^{-|\psi|^2}|\psi\rangle\langle \psi|$. Writing the Hamiltonian in normal order \begin{equation}
\hat H = H_1(\had,\ha)=-\mu\,\had\ha+\frac{U}{2}\had{}^2\ha{}^2,
\end{equation}
one then obtains Eq.~\eqref{eq_BH} using  $\langle \psi|H_1(\had,\ha)|\psi'\rangle = \langle \psi|\psi'\rangle\, H_1(\psi,\psi')$.
By construction, $Z_{1,N}$ is correct to order $\epsilon^2$, and all manipulations made on this expression should not introduce errors of order $\epsilon$.
One then introduces formally the corresponding continuous-time CSPI,
\begin{equation}
\begin{split}
Z_1&=\lim_{N\to \infty}Z_{1,N}=\int\mathcal{D}\psi\mathcal{D}\psi^* e^{-S_1},\\
S_1&=\int_0^\beta d\tau\big( \psi^*(\tau)\partial_\tau\psi(\tau)+H_1\left(\psi^*(\tau),\psi(\tau)\right)\big),
\end{split}
\label{eq_Zcont}
\end{equation}
with periodic boundary conditions $\psi^{(*)}(0)=\psi^{(*)}(\beta)$. Here, and in the following, all ``continuous measures''  are not considered as continuous but actually stand for a shorthand notation for discretized measures.

This path integral is formulated in terms of complex-valued fields, and not in terms of operator-valued fields. Thus, one trades off non-commutativity of operators for time-ordering of fields. And while time-ordering is explicit by construction in the discretized CSPI, it fades away in the continuous limit as the difference between a time $\tau$ and $\tau \pm \epsilon $ becomes ambiguous.
 Fortunately, this ambiguity is easy to solve in perturbation theory, where one just has to remember that $\psi^*(\tau)$ always appears at a slightly later time than $\psi(\tau)$ in the Hamiltonian. This leads to the infamous $e^{i\omega_n 0^+}$ convergence factor that one should add when performing formally divergent Matsubara sums in perturbation theory \cite{AltlandBook2010}.

\subsection{Hubbard-Stratonovich transformation and exact evaluation of the partition function}

The partition function $Z_1$ can be computed exactly using a HS transformation, that, as we will see, has its share of subtlety in the continuous-time path integral. Nevertheless, by handling it properly, we will recover the exact partition function of the Bose-Hubbard model.

The interaction part of $H_1(\psi^*(\tau),\psi(\tau))=\frac{U}{2}\psi^*(\tau)^2\psi(\tau)^2$ can be decoupled in the path integral using the identity
\begin{equation}
e^{-\frac{U}{2}\int_0^\beta d\tau\psi^*(\tau)^2\psi(\tau)^2}=\int \mathcal D \rho \,e^{-\int_0^\beta d\tau \left(\frac{\rho(\tau)^2}{2U}-i\rho(\tau)\psi^*(\tau)\psi(\tau)\right)},
\label{eq_HS}
\end{equation}
where the normalization $[{\rm Det}(\delta(\tau-\tau')/U)]^{\frac12}$ of the Gaussian integral over $\rho$ is included in the measure $\mathcal D \rho$ (see below for the discretized version, which defines this manipulation). This expression is not ambiguous in the continuous limit (as long as we keep in mind the implicit time ordering  of $\psi^*(\tau)\psi(\tau)$), but it is important to note right away that 
\begin{equation}
\int \mathcal D \rho \,\rho(\tau_1)\rho(\tau_2)e^{-\int_0^\beta d\tau \frac{\rho(\tau)^2}{2U}}=U \delta(\tau_1-\tau_2),
\label{eq_flucrho}
\end{equation}
meaning that $\rho(\tau)$ is delta-correlated, much like the white noise of a stochastic process.

We can write the partition function in terms of two functional integrals 
\begin{equation}
\begin{split}
Z_1&=\int \mathcal D \rho \mathcal{D}\psi\mathcal{D}\psi^*\, e^{-S_\rho-S_\psi },\\
S_\rho &= \int_0^\beta d\tau \frac{\rho(\tau)^2}{2U},\\
S_\psi &= \int_0^\beta d\tau\Big(\psi^*(\tau)\partial_\tau\psi(\tau)-(\mu+i\rho(\tau))\psi^*(\tau)\psi(\tau)\Big),
\label{eq_psirho}
\end{split}
\end{equation}
and the strategy is now to perform the integral over the bosons first,\footnote{For positive chemical potential, one should deform the contour on which $\rho$ is integrated over to insure convergence.} to obtain an effective action for the field $\rho$, that we can then evaluate exactly. Since the integral over $\psi$ and $\psi^*$ is gaussian, it gives a functional determinant
\begin{equation}
\int\mathcal{D}\psi\mathcal{D}\psi^*\, e^{-\int_0^\beta d\tau\int_0^\beta d\tau' \psi^*(\tau) F[\tau,\tau';i\rho]\psi(\tau')}={\rm Det}(F[i\rho])^{-1},
\end{equation}
with $F[\tau,\tau';i\rho]=\delta(\tau-\tau')(\partial_{\tau'}-\mu-i\rho(\tau'))$~\cite{AltlandBook2010}. We stress that the underlying time-ordering of the bosons is crucial to compute this functional determinant. With the normal order used here, the functional determinant of $F[\tau,\tau';\Omega]$ with $\Omega(\tau)$ a smooth function reads
\begin{equation}
{\rm Det}(F[\Omega])=1-e^{\int_0^\beta d\tau (\mu+\Omega(\tau))},
\label{eq_detF}
\end{equation}
which is obtained as follows. The discretized version of $\int_0^\beta d\tau\int_0^\beta d\tau' \psi^*(\tau) F[\tau,\tau';\Omega]\psi(\tau')$ is rewritten as $\sum_{k=1}^N\psi^*_k F_{k,k'}[\Omega]\psi_{k'}$ with
\begin{equation}
F_{k,k'}[\Omega]=\delta_{k,k'}-\delta_{k-1,k'}\left(1+\epsilon\m+\epsilon \Omega_k\right).
\end{equation}
with $\tau=k \epsilon$, $\tau'=k' \epsilon$ and $\Omega_k=\Omega((k+\alpha)\epsilon)$, $-1\leq\alpha\leq 0$.\footnote{Since $\Omega(\tau)$ is smooth, changing the point of discretization gives corrections of order $\epsilon^2$, which vanish in the continuous limit.} An explicit calculation of the determinant of $F_{k,k'}[\Omega]$ gives 
\begin{equation}
\begin{split}
\det(F[\Omega])&=1-\prod_{k=1}^N (1+\epsilon\m+\epsilon \Omega_k),\\
&=1-e^{\epsilon\sum_{k=1}^N(\mu+\Omega_k)}+\mathcal{O}(\epsilon^2),
\end{split}
\label{eq_detFd}
\end{equation}
which converges to the functional determinant of Eq.~\eqref{eq_detF} in the continuous limit. Assuming that this formula is valid for the HS field $\rho$, one could try to compute the partition function $Z_1$ as follows,
\begin{equation}
\begin{split}
Z_1&\stackrel{?}{=}\int \mathcal D \rho\,  \frac{e^{-S_\rho}}{1-e^{\beta \mu+i\int_0^\beta d\tau \rho(\tau)}},\\
& =\sum_{n=0}^\infty e^{\beta \mu n}\int\mathcal D \rho\, e^{-\int_0^\beta d\tau\left(\frac{\rho(\tau)^2}{2U}-i n \rho(\tau)\right)},\\
&=\sum_{n=0}^\infty e^{-\beta \left(-\mu n +\frac{U}{2}n^2\right)},
\end{split}
\end{equation}
which does not give the expected result, Eq.~\eqref{eq_exactZ}.

The error comes from the fact that $\rho(\tau)$ is not a smooth field, since its fluctuations diverge at equal time, Eq.~\eqref{eq_flucrho}, and that one cannot exponentiate naively the product in Eq.~\eqref{eq_detFd} in presence of a stochastic field. The discretized versions of Eqs.~\eqref{eq_HS} and \eqref{eq_flucrho} read respectively
\begin{equation}
e^{-\frac{U}{2}\epsilon\sum_{k=1}^N\left(\psi^*_k\psi_{k-1}\right)^2}=\prod_{k=1}^N\int  \frac{d\rho_k}{\sqrt{\frac{2\pi U}{\epsilon}}}e^{-\frac{\epsilon}{2U}\rho_k^2+i\epsilon\rho_k\psi^*_k\psi_{k-1} },
\label{eq_HSdiscrete}
\end{equation}
and 
\begin{equation}
\int \prod_{k=1}^N \frac{d\rho_k}{\sqrt{\frac{2\pi U}{\epsilon}}} \rho_k\rho_{k'}e^{-\frac{\epsilon}{2U}\sum_{k=1}^N\rho_k^2} =\delta_{k,k'}\frac{U}{\epsilon}.
\end{equation}
We should thus think of $\rho_k$ as being of order $\epsilon^{-\frac{1}{2}}$ in all expressions involving it (since it is always integrated over at the end), similarly to the white noise of  (discretized) stochastic processes. Put another way, all physical quantities are averaged over realizations of the white noise $\rho_k$, with a Gaussian distribution of width $1/\epsilon$.

Therefore, when computing the determinant of $F_{k,k'}[i\rho]$, it is \emph{not} true that $(1+\epsilon\m+i\epsilon \rho_k)=e^{\epsilon\m+i\epsilon \rho_k}+\mathcal{O}(\epsilon^2)$. Indeed, the expansion of the exponential gives rise to a term proportional to $\epsilon^2\rho_k^2$ which is in fact of order $\epsilon$, and not $\epsilon^2$ as expected for smooth functions. One therefore needs to correct the exponentiation for stochastic fields, $(1+\epsilon\m+i\epsilon \rho_k)=e^{\epsilon\m+i\epsilon \rho_k+\eps^2\frac{\rho_k^2}{2}}+\mathcal{O}(\epsilon^2)$, which implies
\begin{equation}
\prod_{k=1}^N(1+\epsilon\m+i\epsilon \rho_k)=e^{\sum_{k=1}^N\left(\epsilon\m+i\epsilon \rho_k+\eps^2\frac{\rho_k^2}{2}\right)}+\mathcal{O}(\epsilon^2).
\label{eq_det}
\end{equation}
This expression, while being now correct to order $\epsilon^2$, has the inconvenience to not have a nice continuous limit.

 The proper way to handle stochastic fields is well understood in the context of stochastic (or It\^ o) calculus~\cite{Oksendal2010}. All expressions that are manipulated (such as Eq.\eqref{eq_det}) are to be ``averaged'' over $\rho_k$, and all these $\rho_k$ are independent and of variance $\frac{U}{\epsilon}$. It can then be shown that replacing $\eps^2\frac{\rho_k^2}{2}$ by $\eps\frac{U}{2}$ in all these expressions give a vanishing error in the limit $\epsilon\to 0$ (in a sense that can be made rigorous~\cite{Oksendal2010}). This is the so-called It\^ o's substitution rule (which can also be shown explicitly by computing the average over $\rho_k$ of all expressions with or without the substitution, and showing that the error vanishes with $\epsilon\to 0$), which has the advantage to give expressions with a nice continuous limit, since $\sum_{k=1}^N\eps\frac{U}{2}\to \int_0^\beta d\tau\, \frac{U}{2}$, whereas $\sum_{k=1}^N\eps^2\frac{\rho_k^2}{2}$ does not have one.

 Therefore, in the context of a HS field, the functional determinant of $F[\tau,\tau';\rho]$ is 
 \begin{equation}
 {\rm Det}(F[i\rho])=1-e^{\int_0^\beta d\tau\left(\mu+\frac{U}{2}+i \rho(\tau)\right)},
 \end{equation}
where the correction $\frac{U}{2}$ is similar to a shift of the chemical potential, the origin of which is now understood: it is due to the stochastic nature of the HS field. Using the correct expression for the functional determinant, the remaining integral over $\rho$ now gives the correct result, $Z_1=Z$. 

We note that in Ref.~\onlinecite{Taniguchi2017}, the question of the subtlety of the Hubbard-Stratonovich transformation is raised, and it is proposed to perform a shift of chemical potential $\mu\to\mu+\frac{U}{2}$ directly at the level of the right-hand side of Eqs.~\eqref{eq_psirho} and~\eqref{eq_HSdiscrete}. This cannot be so, since the Hubbard-Stratonovich transformation is well defined there, both in its discretized version and in the continuous limit. Indeed, integrating out the HS field before the bosons, one should recover the left-hand side of Eqs.~\eqref{eq_psirho} and~\eqref{eq_HSdiscrete}, and this would not be true if we had arbitrarily shifted the chemical potential. Correction terms, coming from It\^o's substitution, should only appear \emph{after} computing the Green's function of the field $\psi$ in presence of $\rho$, or in functional determinants (in fact, this is how the presence of this correction term was understood in Appendix C of Ref.~\onlinecite{Taniguchi2017}). Just shifting the chemical potential in Eq.~\eqref{eq_psirho}, and then forgetting about  It\^o's correction term when computing functional determinants is an inconsistent way of performing the path integral evaluation. The same It\^o prescription should be used in the Lie-Algebraic approach to the HS transformation of Ref.~\onlinecite{Galitski2011}, as well as in the calculations of Ref.~\onlinecite{Kordas2019} Section III, where it is missing.

\section{Arbitrary ordering and path integral correspondence\label{ordering} }

We now address the question of ordering and its relation to the discretization scheme, which links the time-slices on which the fields $\psi$ and $\psi^*$ live  in the CSPI, and the form of the corresponding action $S$. The only assumption is that the Hamiltonian operator $\hat H$, if written in normal order $H_1(\hat a^\dag,\hat a)$, is of the form
\begin{equation}
H_1(\hat a^\dag,\hat a) = \sum_q g_q \hat a^\dag {}^q \hat a^q,
\end{equation}
which for the single-site Bose-Hubbard model corresponds to $g_0=0$, $g_1=-\mu$ and $g_2=\frac{U}{2}$ ($g_{q>2}=0$). We stress that $\hat H$ and $H_1(\hat a^\dag,\hat a)$ represent the same operator, with identical spectrum and eigenstates. It is only the way of writing them that can be different. As discussed previously, the corresponding discretized action of the (normal ordered) CSPI is
\begin{equation}
S_{1}=\sum_{k=1}^N \left(\psi_k^*(\psi_k-\psi_{k-1})+\epsilon H_1(\psi_k^*,\psi_{k-1})\right).
\end{equation}
Assume now that the CSPI can be expressed in an exact fashion with a different discretization
\begin{equation}
S_{s}=\sum_{k=1}^N \left(\psi_k^*(\psi_k-\psi_{k-1})+\epsilon H_s(\psi_k^*,\psi_{k_s})\right),
\label{eq_Ss}
\end{equation}
where
\begin{equation}
\psi_{k_s}=\frac{1-s}{2}\psi_k+\frac{1+s}{2}\psi_{k-1},
\end{equation}
and $H_s$ is the corresponding function for this discretization scheme $s$. 
 Obviously, for $s=1$, we recover the standard discretization scheme (normal ordered), and $H_1(\psi_k^*,\psi_{k-1})$ is the function obtained from the operator $\hat H$ when written in normal order, followed by the  formal correspondence  $\hat a^\dag \to \psi^*_k$,  $\hat a \to \psi_{k-1}$.
 One also shows that for a Gaussian action, $\langle\psi_k^*\psi_{k_s}\rangle_G=\frac{1-s}{2}\langle\hat a \hat a^\dag\rangle_G+\frac{1+s}{2}\langle\hat a^\dag \hat a\rangle_G$ and thus terms like $\psi_k^*\psi_{k_s}$ should be understood as coming from $\frac{1-s}{2}\hat a \hat a^\dag+\frac{1+s}{2}\hat a^\dag \hat a$. Then $s=-1$ would correspond to anti-normal order, while $s=0$ is the Weyl (symmetric) ordering. It has been speculated quite some time ago that a change of discretization point in the Hamiltonian symbol $H_s$ is related to a problem of operator ordering of the Hamiltonian $\hat H$~\cite[p. 262]{SchulmanBook}. This is confirmed in the specific case $s=-1$ when constructing the CSPI using the P-representation of the Hamiltonian~\cite{Marchioro1990,Santos2006,Bruckmann2018}. We now address the case of an arbitrary discretization $s$. Using only transformations  consistent with the rules of stochastic calculus in path integrals,  we relate the function $H_s$ to $H_1$, or equivalently to the ordering of the operator $\hat H$, and we find the appropriate correspondence between operators and fields.

We start by rewriting a typical term of $H_1$ as
\begin{equation}
\begin{split}
(\psi_k^*\psi_{k-1})^q&=\left(\psi_k^*\left(\psi_{k_s}+\frac{s-1}{2}\Delta\psi_k\right)\right)^q,\\
&=\sum_{p=0}^q \binom{q}{p}\left(\frac{s-1}{2}\right)^p\left(\psi_k^*\Delta\psi_k\right)^p \left(\psi_k^*\psi_{k_s}\right)^{q-p},
\end{split}
\end{equation} 
with $\Delta\psi_k=\psi_k-\psi_{k-1}$. The action now reads
\begin{equation}
S=\sum_{k=1}^N \left(\psi_k^*\Delta\psi_k+\epsilon \sum_q g_q \sum_{p=0}^q \binom{q}{p}\left(\frac{s-1}{2}\right)^p\left(\psi_k^*\Delta\psi_k\right)^p \left(\psi_k^*\psi_{k_s}\right)^{q-p}\right).
\label{eq_Sint}
\end{equation}
Naively taking the continuous limit amounts to replace $\Delta\psi_k\to \epsilon\,\partial_\tau \psi(\tau=k\epsilon)$, which removes all terms with $p>0$. However, this would imply that $H_s(\psi_k^*,\psi_{k_s})=H_1(\psi_k^*,\psi_{k_s})$, which cannot be true. We cannot recover the perturbative calculation of $Z$ using the same function $H_1$, but a different ordering rule for $\langle\psi(\tau)\psi(\tau)\rangle$.
We, therefore, have to be careful with the handling of the $p>0$ terms, which cannot be discarded too quickly.

In both the cases of Feynman's and stochastic path integrals, the way to handle extra terms that should naively vanish but that nevertheless contribute to the action is known~\cite{McLaughlin1971,Cugliandolo2017}. There, the kinetic term is of the form $\Delta x_k^2/\epsilon$, where $x_k$ is the position at time-slice $k$, and $\Delta x_k=x_k-x_{k-1}$. This implies that $\Delta x_k$ is of order $\sqrt{\epsilon}$, while extra terms are typically of the form $\Delta x_k^3/\epsilon$ or $\Delta x_k^4/\epsilon$, which contribute to order $\sqrt\epsilon$ and $\epsilon$ respectively. Expanding $e^{-S}$ to order $\epsilon$, and using replacement rules equivalent to that of It\^o, one then rewrites the action in a way that allows for a simple continuous limit. The crucial steps are to correctly figure out the replacement rules to use, to correctly expand $e^{-S}$, and then re-exponentiate.

In our case, the replacement rules cannot be as simple, because the ``kinetic'' term, $\psi_k^*\Delta\psi_k$, is of order $\epsilon^0$ (since it is not multiplied by any power of $\epsilon$). Therefore, all terms $\left(\psi_k^*\Delta\psi_k\right)^p$ will contribute in the continuous limit. However since $\left(\psi_k^*\Delta\psi_k\right)^p \left(\psi_k^*\psi_{k_s}\right)^{q-p}$ is multiplied by a factor $\epsilon$ in Eq.~\eqref{eq_Sint},  it is sufficient  to find the replacement rules of $\left(\psi_k^*\Delta\psi_k\right)^p$ to order $\epsilon^0$, as additional terms in $\epsilon$ would give corrections of order $\epsilon^2$ in total. To uncover the replacement rules, it is thus convenient to introduce the generating function
\begin{equation}
f_s(x,y)=\int \prod_{k=1}^{N}\frac{d\psi_k d\psi_k^*}{2i\pi} e^{-S_f(x,y)},
\label{eq_fs}
\end{equation}
with 
\begin{equation}
S_f(x,y)=\sum_{k=1}^N \left((1+\delta_{kl}x)\psi_k^*(\psi_k-\psi_{k-1})+(\epsilon a+\delta_{kl}y)\psi_k^*\psi_{k_s}\right),
\label{eq_Sf}
\end{equation} 
with $a>0$ needed to insure convergence. Additional terms of order $\epsilon$ in $S_f$ would give irrelevant corrections in the limit $\epsilon\to0$. An explicit calculation gives
\begin{equation}
f_s(x,y)=\frac{1}{A+B x+C y},
\end{equation}
with
\begin{equation}
\begin{split}
A&=\left(1+\frac{1-s}{2}a\epsilon\right)^N-\left(1-\frac{1+s}{2}a\epsilon\right)^N,\\
B&=\left(1+\frac{1-s}{2}a\epsilon\right)^{N-1}-\left(1-\frac{1+s}{2}a\epsilon\right)^{N-1},\\
C&=\frac{1-s}{2}\left(1+\frac{1-s}{2}a\epsilon\right)^{N-1}+\frac{1+s}{2}\left(1-\frac{1+s}{2}a\epsilon\right)^{N-1}.
\end{split}
\end{equation} 
This implies that
\begin{equation}
\begin{split}
\langle\left(\psi_l^*\Delta\psi_l \right)^p \left(\psi_l^*\psi_{l_s}\right)^{q-p}\rangle&=\frac{(-1)^{q}}{f_s(0,0)}\frac{\partial^{q}f_s}{\partial x^p\partial y^{q-p}}\bigg|_{x=y=0},\\
&=q! \frac{B^p C^{q-p}}{A^{q}},\\
&=q! \frac{C^{q-p}}{A^{q-p}}+\mathcal{O}(\epsilon),\\
&=\frac{q!}{(q-p)!}\langle\left(\psi_l^*\psi_{l_s}\right)^{q-p}\rangle+\mathcal{O}(\epsilon),
\end{split}
\end{equation}
since $B=A+\mathcal{O}(\epsilon)$. We therefore find the non-trivial replacement rule, valid to order $\epsilon$, and to be used directly into Eq.~\eqref{eq_Sint} to the same precision,
\begin{equation}
\left(\psi_k^*\Delta\psi_k \right)^p \left(\psi_k^*\psi_{k_s}\right)^{q-p}\to p!\binom{q}{p}\left(\psi_k^*\psi_{k_s}\right)^{q-p}.
\end{equation}

The action is then finally written in a way that allows for a simple continuous limit, with discretization $s$, given by Eq.~\eqref{eq_Ss}, with the function
\begin{equation}
H_s(\psi^*_k,\psi_{k_s})= \sum_q g_q \sum_{p=0}^q p!\binom{q}{p}^2\left(\frac{s-1}{2}\right)^p\left(\psi_k^*\psi_{k_s}\right)^{q-p}.
\label{eq_Hs}
\end{equation}
Interestingly, this form is closely related to the rule converting normal-ordered products to general $s$-ordered products of bosonic operators, as introduced by Cahill and Glauber~\cite{Cahill1969},  which reads
\begin{equation}
\hat a^\dag {}^q \hat a^q=\sum_{p=0}^q p!\binom{q}{p}^2\left(\frac{s-1}{2}\right)^p \left\{(\hat a^\dag\hat a)^{q-p}\right\}_s,
\end{equation}
with $\{(\hat a^\dag\hat a)^n\}_s$ the $s$-ordered product of $(\hat a^\dag\hat a)^n$. In this formalism, $s=1,0$, and $-1$ corresponds respectively to normal, Weyl, and anti-normal ordering, although continuous ordering indices $s\in[-1,1]$ are also defined. In particular, $\{\hat a^\dag\hat a\}_s=\frac{1-s}{2}\hat a \hat a^\dag+\frac{1+s}{2}\hat a^\dag \hat a$ for all $s\in[-1,1]$. 

We thus conclude that the continuous-time limit CSPI is consistent with arbitrary $s$-ordering  of the Hamiltonian, with action
\begin{equation}
S_s=\int_0^\beta d\tau \Big(\psi^*(\tau)\partial_\tau \psi(\tau)+H_s(\psi^*(\tau),\psi(\tau))\Big),
\end{equation}
if interpreted as follow: i) $H_s(\psi^*(\tau),\psi(\tau))$ is obtained from the operator $\hat H$ by taking its  $s$-ordering $\{\hat H\}_s$ and replacing $\hat a^\dag\to \psi^*(\tau)$ and $\hat a\to \psi(\tau)$; ii) when ambiguous, $\psi^*(\tau) \psi(\tau)$ should be interpreted as the continuous limit of $\psi_k^*\psi_{k_s}$, i.e. as $\frac{1-s}{2}\psi(\tau+0^+) \psi^*(\tau)+\frac{1+s}{2}\psi^*(\tau+0^+) \psi(\tau)$.\footnote{This amounts to add a convergence factor $\frac{1-s}{2}e^{-i\omega_n 0^+}+\frac{1+s}{2}e^{i\omega_n 0^+}$ to divergent Matusbara sums.} Eq.~\eqref{eq_Hs} is in agreement with the anti-normal ordered CSPI used in Refs.~\onlinecite{Santos2006,Bruckmann2018}, but is rather different from the Weyl-ordered CSPI defined in Ref.~\onlinecite{Santos2006}. Calculations  for arbitrary $s$  similar to that of Sec.~\ref{exactBH} gives the exact partition function of the single-site Bose-Hubbard model.~\footnote{For arbitrary $s$-ordering, the functional determinant Eq.~\eqref{eq_detF} should be generalized to ${\rm Det}(F[\Omega])_s=e^{-\frac{1-s}{2}\int_0^\beta d\tau (\mu+\Omega(\tau))}-e^{\frac{1+s}{2}\int_0^\beta d\tau (\mu+\Omega(\tau))}$. This can be shown by taking the continuous limit of the determinant obtained in the corresponding discretization scheme $s$, similar to the calculation of $f_s(x=0,y=0)=1/A$. If $\Omega$ is stochastic, correction terms should be added using  It\^o's substitution rule.}

\section{Discussion \label{discuss}}

The calculation of path integrals is a subtle problem, even for the simplest models. ``Exact calculations'' are as valid as the manipulations made to perform them, and unfortunately, the standard rules of calculus, such as the chain rule in non-linear changes of variables, or even solving differential equations, do not necessarily work the same way in path integrals.
 In particular, we have shown that in the case of CSPI, the substitution rules are more involved than in stochastic and Feynman path integrals, because the difference between two time-slices $\Delta \psi_k$ is of order $\epsilon^0$, and not of order $\sqrt\epsilon$ as usual. This has important consequences in the case of non-linear change of variables.

In Ref.~\onlinecite{Wilson2011}, Wilson and Galitski use the amplitude-phase representation, which amounts to do the (well-defined) change of variable $\psi_k^{(*)}=\sqrt{\rho_k}e^{(-)i\theta_k}$, with unit Jacobian. Then, assuming that one can replace $\psi^*(\tau)\partial_\tau \psi(\tau)$ by $\frac{1}{2}\partial_\tau \rho(\tau)+i\rho(\tau) \partial_\tau \theta(\tau)$ in the continuous limit, they could not recover the correct partition function of the single-site Bose-Hubbard model. The change of variable at the level of the discrete path integral is perfectly valid, and performing the calculation there does indeed give the correct result, as shown by Bruckmann and Urbina~\cite{Bruckmann2018}. The problem lies in taking the continuous limit, in the replacement
\begin{equation}
\psi_k^*(\psi_k-\psi_{k-1})=\rho_k-\sqrt{\rho_k\rho_{k-1}}e^{-i(\theta_k-\theta_{k-1})}\to \frac{\rho_k-\rho_{k-1}}{2}+i\rho_k(\theta_k-\theta_{k-1}).
\end{equation}
Indeed, in this term, both $\rho_k-\rho_{k-1}$ and $\theta_k-\theta_{k-1}$ are not multiplied by any power of $\epsilon$, and thus appear to be of order $\epsilon^0$. As discussed previously, there is no reason to neglect higher powers of $\rho_k-\rho_{k-1}$ and $\theta_k-\theta_{k-1}$, which need to be taken care of. This requires to develop consistent replacement rules beyond the ones found here, which does not seem to be a simple task. And while Bruckmann and Urbina have managed to write an exact form of the continuous-time path integral in the amplitude-phase representation, it is based on a series of dual transformations that do not seem to be suited to other non-linear changes of variables. It would also be very interesting to find a formulation of the CSPI that allows for naive change of variable (i.e. using the standard rules of calculus), as was recently found for stochastic path integrals~\cite{Cugliandolo2018}. 

One should also investigate the implications of the stochastic nature of Hubbard-Stratonovich fields uncovered here. In particular, the functional determinants, obtained after integrating out the bosons (or fermions) decoupled by the HS transformation, are rarely calculated exactly. Instead, mean-field approximations are made on the HS field, which therefore loses its stochastic nature, and the corresponding correction term seems to disappear. Due to the prevalence of this method in condensed matter theory, it is important to investigate when and where these correction terms should appear.

Another open question in the context of CSPI is the fate of semi-classical approximations. There are still controversies on which ordering of the Hamiltonian should be used  to get the correct semi-classical calculation since quite surprisingly, the different orderings do not give the same semi-classical result~\cite{Santos2006,Wilson2011}. Whether the aspects discussed in this manuscript, both on ordering and discretization, and on the stochasticity of the fields, could shine a new light on this issue deserves to be explored.

\subsection*{ACKNOWLEDGMENTS}
We thank  B. Arras and J. D. Urbina for discussions, E. Kochetov for comments, and N. Dupuis for a careful reading of the manuscript. This study was supported by the French government through the Programme Investissement d'Avenir (I-SITE ULNE/ANR-16-IDEX-0004 ULNE) managed by the Agence Nationale de la Recherche.

%
%
%
%
%


\end{document}